\documentclass[a4paper]{jpconf}
\usepackage{graphicx}
\usepackage{psfrag}
\begin{document}
\begin{flushright}
NORDITA-2010-29, UUITP-15/10
\end{flushright}

\title{Holographic models with anisotropic scaling}

\author{E J Brynjolfsson$^1$, U H Danielsson$^2$, L Thorlacius$^{1,3}$ and
T Zingg$^{1,3}$}

\address{$^1$ University of Iceland, Science Institute, Dunhaga 3, IS-107 Reykjavik, Iceland}
\address{$^2$ Institutionen f\"or fysik och astronomi, Uppsala Universitet, Box 803,
SE-751 08 Uppsala, Sweden}
\address{$^3$ NORDITA, Roslagstullsbacken 23, SE-106 91 Stockholm, Sweden}

\ead{erlingbr@hi.is ulf.danielsson@physics.uu.se lth@nordita.org zingg@nordita.org}

\begin{abstract}
We consider gravity duals to d+1 dimensional quantum critical points with 
anisotropic scaling. The primary motivation comes from strongly correlated
electron systems in condensed matter theory but the main focus of the 
present paper is on the gravity models in their own right. Physics at finite
temperature and fixed charge density is described in terms of charged
black branes. Some exact solutions are known and can be used to obtain
a maximally extended spacetime geometry, which has a null curvature 
singularity inside a single non-degenerate horizon, but generic black brane 
solutions in the model can only be obtained numerically. Charged matter 
gives rise to black branes with hair that are dual to the superconducting 
phase of a holographic superconductor. Our numerical results indicate 
that holographic superconductors with anisotropic scaling have vanishing 
zero temperature entropy when the back reaction of the hair on the
brane geometry is taken into account.
\end{abstract}

\section{Introduction}
We consider $d+2$ dimensional gravity models that are dual to quantum critical
points with anisotropic scaling in $d+1$ dimensions, 
\begin{equation}
t\rightarrow\lambda^{z}t,\qquad \vec{x}\rightarrow\lambda \vec{x}\,,
\label{lscaling}
\end{equation}
with $z\geq 1$ and $\vec{x}=(x_1,\ldots,x_d)$. 
The $d+2$ dimensional bulk action consists of three parts,
\begin{equation}
S  = S_\textrm{\small Einstein-Maxwell} + S_\textrm{\small Lifshitz}
+S_\textrm{\small matter} \,.
\label{action}
\end{equation}
The first term is the standard action of Einstein-Maxwell gravity with a 
negative cosmological constant,
\begin{equation}
S_\textrm{\small Einstein-Maxwell} = \int\mathrm{d}^{d+2}x\sqrt{-g}\;\left(  
R-2\Lambda-\frac{1}{4}F_{\mu\nu}F^{\mu\nu}\right) .
\label{einsteinmaxwell}
\end{equation}
This is followed by a term involving a massive vector field,
\begin{equation}
S_\textrm{\small Lifshitz} = - \int\mathrm{d}^{d+2}x\sqrt{-g}\;
\left(\frac{1}{4} \mathcal{F}_{\mu\nu}\mathcal{F}^{\mu\nu} 
+\frac{c^2}{2}\mathcal{A}_\mu\mathcal{A}^\mu\right) ,
\label{lifshitzaction}
\end{equation}
whose sole purpose is to provide backgrounds with anisotropic scaling. 
This is a modified version of the holographic model 
of \cite{Kachru:2008yh}, which was formulated in four-dimensional spacetime
and obtained anisotropic scaling by including a pair of coupled two- and
three-form field strengths.\footnote{See also \cite{Koroteev:2007yp} for early work 
on gravitational backgrounds with anisotropic scaling.}
Upon integrating out the three-form field strength, 
the remaining two-form becomes a field strength of a massive vector \cite{Taylor:2008tg} 
and in this form the model is easily extended to general dimensions. Finite temperature
corresponds to having a black hole in the higher-dimensional spacetime and with
the Maxwell gauge field, added in \cite{Brynjolfsson:2009ct}, the black hole can 
carry electric charge, which is dual to a finite charge density in the lower-dimensional
theory. Finally, we consider matter in the form of a scalar field, which is charged under 
the Maxwell gauge field but does not couple directly to the auxiliary massive vector field,
\begin{equation}
S_\textrm{\small matter} = -\frac{1}{2} \int\mathrm{d}^{d+2}x\sqrt{-g}\;\left( 
\left(\partial^\mu\phi^*+iqA^\mu\phi^*\right)
\left(\partial_\mu\phi-iqA_\mu\phi\right)+m^2\phi^*\phi\right) .
\label{matteraction}
\end{equation}
At low temperature there is an instability for charged black holes in the 
model to grow scalar hair, which corresponds to the superconducting phase
of holographic superconductors with $z>1$ asymmetric scaling 
\cite{Brynjolfsson:2009ct}, as described in Section~\ref{holosc} below.

The equations of motion obtained from the action (\ref{action}) consist of the 
scalar field equation 
\begin{equation}
\left(\nabla^\mu-iqA^\mu\right)\left(\nabla_\mu-iqA_\mu\right)\phi-m^2\phi=0\,,
\label{mattereom}
\end{equation}
along with the Einstein equations, the Maxwell equations, and field equations for the 
auxiliary massive vector,
\begin{eqnarray}
\label{einsteineqs}
G_{\mu\nu}+\Lambda g_{\mu\nu}&=&T_{\mu\nu}^\textrm{\small Maxwell}
+T_{\mu\nu}^\textrm{\small Lifshitz}+T_{\mu\nu}^\textrm{\small matter} \,, \\
\label{maxwelleqs}
\nabla_\nu F^{\nu\mu} &=&j^\mu_\textrm{\small matter} \,, \\
\label{lifshitzeqs}
\nabla_\nu \mathcal{F}^{\nu\mu} &=& c^2 \mathcal{A}^\mu \,.
\end{eqnarray}
The asymmetric scaling symmetry (\ref{lscaling}), sometimes referred to as Lifshitz scaling, 
is realized in a d+2 dimensional spacetime,
\begin{equation}
ds^2=L^2\left(  -r^{2z}dt^2+r^2d\vec{x}^2+\frac{dr^2}{r^2}\right)  ,\label{lmetric}
\end{equation}
whose metric is invariant under the transformation
\begin{equation}
t\rightarrow\lambda^{z}t,\quad 
\vec{x}\rightarrow\lambda \vec{x}, \quad
r\rightarrow\frac{r}{\lambda}\,.
\label{lifshitzscaling}
\end{equation}
Length dimensions are carried by the characteristic length $L$ while the 
coordinates $(t,r,\vec{x})$ are dimensionless. The scale invariant Lifshitz geometry 
(\ref{lmetric}) is a solution of the equations of motion 
when $L$ is related to the cosmological constant 
$\Lambda$ via
\begin{equation}
\Lambda=-\frac{z^{2}+(d{-}1)z+d^2}{2L^{2}}\,,
\label{cosmoconstant}
\end{equation}
and the mass of the auxiliary vector field is fine-tuned to $c=\sqrt{z\,d}/L$. 
In this solution the Maxwell field vanishes, $A_\mu=0$, but the massive vector field 
has the background value
\begin{equation}
\mathcal{A}_t=\sqrt{\frac{2(z{-}1)}{z}}L\,r^z \,, \qquad 
\mathcal{A}_{x_i}=\mathcal{A}_r=0 \,.
\label{lbackground}
\end{equation}

\section{Charged black branes }
In order to study finite temperature effects in the dual strongly coupled field theory we 
look for static black brane solutions of the equations of motion 
(\ref{mattereom}) - (\ref{lifshitzeqs}) which are asymptotic to the Lifshitz fixed point
solution given by (\ref{lmetric}) and (\ref{lbackground}). From now on we set $L=1$ 
and consider a metric of the form 
\begin{equation}
ds^2 =-r^{2z}f(r)^2 dt^2+r^2d\vec{x}^2 +\frac{g(r)^2}{r^2}dr^2 \,.
\label{metricansatz}
\end{equation}
An asymptotically Lifshitz black brane with a non-degenerate horizon
has a simple zero of both $f(r)^2$ and $g(r)^{-2}$ at the horizon, which we take to be 
at $r=r_0$, and $f(r),\,g(r) \rightarrow 1$ as $r\rightarrow\infty$. 
It is straightforward to generalize this ansatz to include black holes with a spherical 
horizon or topological black holes with a hyperbolic horizon but it is the flat horizon 
case (\ref{metricansatz}) that is of direct interest for the gravitational dual description 
of strongly coupled $d+1$ dimensional field theories. In a static electrically charged 
black brane background the Maxwell gauge field and the 
massive vector can be taken to be of the form
\begin{equation}
A_\mu= r^zf(r)\{\alpha(r),0,\ldots,0\} \,; \qquad 
\mathcal{A}_\mu = \sqrt{\frac{2(z{-}1)}{z}}r^zf(r) \{a(r),0,\ldots,0\} \,.
\end{equation}
with $\alpha(r)\rightarrow 0$ and $a(r)\rightarrow 1$ as $r\rightarrow\infty$.

\subsection{Field equations for static configurations}

For static configurations the equations of motion can be expressed as a 
first order system of ordinary differential equations. Introducing a scale invariant
radial variable $u=\log(r/r_0)$ and writing $\dot{\ }\equiv\frac{d}{du}$,
the scalar field equation becomes
\begin{eqnarray}
\dot{\phi}&=&\chi \,, \label{scalar1}\\
\dot{\chi}+ \left(\frac{\dot{f}}{f}-\frac{\dot{g}}{g}\right)\chi&=&-\left(d{+}z\right)\chi+
\left(m^2-q^2\alpha^2\right)g^2\phi\,,
\label{scalar2}
\end{eqnarray}
while the Maxwell equations and the equations
of motion for the massive vector reduce to,
\begin{eqnarray}
\dot{\alpha}+\frac{\dot{f}}{f} \alpha&=& -z\, \alpha+g\, \beta\,, 
\label{maxwell1u} \\
\dot{\beta}&=&-d\, \beta+q^2g\,\phi^2\alpha \,,
\label{maxwell2u} \\
\dot{a}+\frac{\dot{f}}{f}a&=& -z\, a+z\, g\, b \,, 
\label{lifshitz1u} \\
\dot{b}&=& -d\, b+d\, g\, a\,.
\label{lifshitz2u} 
\end{eqnarray}
The functions $\chi$, $\beta$, and $b$ are defined via 
(\ref{scalar1}), (\ref{maxwell1u}), and (\ref{lifshitz1u}), respectively. 
The Einstein equations can also be written in first order form,
\begin{eqnarray}
\frac{\dot{g}}{g}{+}\frac{\dot{f}}{f}&=&  (z{-}1)\left(g^2a^2-1\right)
+\frac{1}{2d}\left(\chi^2+q^2g^2\alpha^2\phi^2\right), 
\label{einstein2} \\
2d\,\frac{\dot{f}}{f}&=& d(1{-}d{-}2z)+\frac{\chi^2}{2}+
g^2\! \left[(z{-}1)\left(da^2-zb^2\right)-\frac{\beta^2}{2} \right. \nonumber \\
&\phantom{=}&\qquad\left.
+\frac{1}{2}\left(q^2\alpha^2-m^2\right)\phi^2+z^{2}+(d{-}1)z+d^2
\right]. \label{einstein1} 
\end{eqnarray}
The field equations (\ref{scalar1}) - (\ref{einstein1}) are manifestly invariant under the 
scaling (\ref{lifshitzscaling}) and the Lifshitz fixed point solution is given by 
$f=g=a=b=1$ and $\alpha=\beta=\phi=\chi=0$. 

In the absence of charged matter, the Maxwell equation (\ref{maxwell2u}) 
integrates to $\beta(u) = \tilde{\rho}\, e^{-d u}$.
The integration constant $\tilde{\rho}=\rho/r_0^d$ is proportional to the electric charge 
per unit $d$-volume of the black brane, which in turn corresponds to the charge density 
in the dual field theoretic system. For given $d$ and $z\geq 1$, 
the remaining field equations then have a one parameter family of black brane solutions, 
labelled by $\tilde{\rho}$. A neutral black brane without scalar hair has $\tilde{\rho}=0$ 
while the extremal limit is given by 
$\tilde{\rho}\rightarrow \pm\sqrt{2(z^{2}+(d{-}1)z+d^2)}$. A non-vanishing charged
scalar field changes this picture as discussed below.

Equation (\ref{maxwell1u}) can easily be
solved in the Lifshitz geometry (\ref{lmetric}) without matter, giving
\begin{equation}
\alpha(u) = \left\{ 
\begin{array}{lcc}
\tilde{\mu}\,e^{-zu}+\frac{1}{(z-d)}\tilde{\rho}\,e^{-du}
& \quad\textrm{if}\quad & z\neq d \,, \\
\tilde{\mu}\,e^{-du}+\tilde{\rho}\,u\,e^{-du}
& \quad\textrm{if}\quad & z= d \,,
\end{array}
\right.
\label{Apotential}
\end{equation}
where the integration constant $\tilde{\mu}=\mu/r_0^z$ corresponds to having 
non-vanishing chemical potential in the dual system. 
In general, we are not working with the Lifshitz background but with
solutions that are only asymptotically Lifshitz as $u\rightarrow\infty$. 
However, as long as the value of $z$ isn't too high,\footnote{At $z\geq 3d$ non-linear 
effects give rise to additional terms in (\ref{Apotential}) with a falloff in between 
that of the charge density and chemical potential terms.} 
the asymptotic behavior of the gauge potential carries over 
from the Lifshitz background to the more general case, and one can read off the charge 
density and chemical potential in the dual field theory from the leading two terms
in the expansion of $\alpha$ at large $u$. Calculations in this paper refer to fixed
$\rho$, corresponding to a fixed density of charge carriers in
the dual system, but one can also work at fixed chemical potential. 

\subsection{Numerical solutions}

Black brane solutions at generic $z> 1$ can be obtained using 
numerical techniques similar to those of \cite{Danielsson:2009gi}. 
The field equations are integrated numerically starting near the
black hole, with suitable initial conditions and proceeding out towards 
the asymptotic region. For a regular, non-degenerate horizon we require
the functions that appear in the metric ansatz (\ref{metricansatz}) to behave as
\begin{equation}
f(u)=\sqrt{u}(f_0+f_1 u+\ldots), \quad  g(u)=\frac{1}{\sqrt{u}}(g_0 +g_1 u+\ldots),
\label{horizonexp}
\end{equation}
near $u=0$. The appropriate near-horizon behavior of the remaining field variables
can be worked out and then inserted into the equations of motion to generate
initial value data for the numerical integration. For any given values of $d$ and $z$, we
obtain a three parameter family of initial values, 
where for instance $\phi(0)$, $\beta(0)$, and $b(0)$ can be taken as the independent 
parameters. The condition that the metric and massive vector approach the 
Lifshitz fixed point solution (\ref{lmetric}) and (\ref{lbackground}) sufficiently 
rapidly as $u\rightarrow \infty$ restricts the solutions 
further \cite{Danielsson:2009gi,Bertoldi:2009vn,Ross:2009ar}. As a result,
$b(0)$ is fixed for given $\phi(0)$ and $\beta(0)$ and one has a two-parameter
family of solutions. This means in particular that, in the absence of scalar hair, 
there is a unique (up to overall scale) asymptotically Lifshitz charged black brane 
solution for given $d$, $z$ and $\tilde{\rho}$.

The Hawking temperature is determined by the near-horizon behavior of the 
black brane metric,
\begin{equation}
T_H=\frac{r_0^z}{4\pi}\,\frac{f_0}{g_0} \,.
\label{hawkingtemp}
\end{equation}
The full numerical solution of the field equations is required, however, to relate the 
coefficients $f_0$ and $g_0$ to the charge density $\rho$ and other physical 
variables of the dual field theory. 

\subsection{Conserved charge under radial evolution}

A conserved charge under radial evolution was found in \cite{Bertoldi:2009vn} for 
electrically neutral black branes with $d=2$ and arbitrary dynamical critical 
exponent $z$. Such a conserved charge is useful for matching solutions
across the bulk geometry from the near-horizon region to the asymptotic large $u$
region and also provides a check on numerical solutions. The charge found in 
\cite{Bertoldi:2009vn} generalizes to charged black branes with scalar hair in 
general spatial dimensions,
\begin{eqnarray}
D_0&=& r_0^{z+d}e^{(z+d)u}f\left[
\frac{1}{2g}(\chi^2-2d(d{+}1))-2d(z{-}1)ab-d\,\alpha\beta \right.\nonumber \\
&\ &\left.\quad +g\left[(z{-}1)(d\, a^2-z\,b^2)+z^2{+}(d{-}1)z{+}d^2
-\frac{1}{2}(\beta^2+m^2\phi^2-q^2\alpha^2\phi^2)\right]\right].
\label{conservedcharge}
\end{eqnarray} 
It is straightforward to check that $\frac{d}{du}D_0=0$ when the field equations
(\ref{scalar1}) - (\ref{einstein1}) are satisfied. The conserved charge is related to
thermodynamic state variables of the dual system in a simple way. 
Inserting a perturbative near-horizon expansion of the fields, one finds
\begin{equation}
D_0= 2r_0^{d+z}\frac{f_0}{g_0} =32\pi S\,T,
\label{nearbh}
\end{equation}
where $T$ is the temperature (\ref{hawkingtemp}) and $S=r_0^d/4$ is the 
Bekenstein-Hawking entropy density, which are identified with the temperature
and entropy of the dual field theory.

\subsection{Exact solution}

As always, it is useful to have explicit analytic solutions to work with. Although
Lifshitz black branes at $z>1$ can in general only be obtained numerically, 
it turns out that for each value of $d$ an isolated $z=2d$ exact solution can be found 
\cite{Brynjolfsson:2009ct,Pang:2009pd},\footnote{The corresponding exact solutions for 
$d=2$, $z = 4$ black holes with a spherical horizon and topological black holes with 
a hyperbolic horizon were also found in \cite{Brynjolfsson:2009ct}. In the limit of 
vanishing electric charge these black hole solutions reduce to the previously 
discovered $z=4$ black hole solution of \cite{Bertoldi:2009vn}.}
\begin{equation}
b=1\,, \quad f^2=\frac{1}{g^2}=a^2=1-e^{-2du}, \quad 
f\alpha=\pm\sqrt{2}e^{-du}\,(1-e^{-du}),
\label{z4exact}
\end{equation}
with $\phi=0$ and $\tilde{\rho}=\pm\sqrt{2}d$. 
It is straightforward to continue the exact black brane metric inside the 
horizon and obtain the globally extended geometry \cite{Brynjolfsson:2009ct}.
Define a tortoise coordinate $u_*$ by
\begin{equation}
u_*=\frac{1}{2d\,r_0^{2d}}\log \left( 1-e^{-2du}\right)\,,
\label{tortoise}
\end{equation}
and then transform the $(t,u)$ variables to a pair of null coordinates
\begin{equation}
V=\exp\left[d\, r_0^{2d}(u_*{+}t)\right],\quad
U=-\exp\left[d\, r_0^{2d}(u_*{-}t)\right].
\label{kruskals}
\end{equation}
In the new coordinate system the metric is given by 
\begin{equation}
ds^2=\frac{-dU\,dV}{d^2(1+UV)^2}
+\frac{r_0^2\,d\vec{x}^2}{(1+UV)^{1/d}} \,,
\label{globalmetric}
\end{equation}
and is manifestly non-singular at the horizon, which is located at $UV=0$.
\psfrag{y}[c]{$r=0$}
\psfrag{x}[c]{$r=\infty$}
\psfrag{z}[c]{$r=r_0$}
\begin{figure}[h]
\includegraphics[width=3cm]{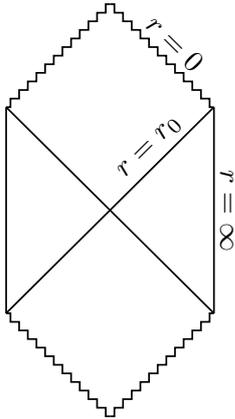}\hspace{2pc}%
\begin{minipage}[b]{20pc}\caption{\label{fig:global}Global geometry of an 
asymptotically Lifshitz black hole. The null singularity at $r=0$ is indicated
by the jagged lines.}
\end{minipage}
\end{figure}
There is a null curvature singularity at $UV\rightarrow\infty$, which corresponds
to $r\rightarrow 0$ in the original coordinate system. The asymptotic region 
$r\rightarrow\infty$ corresponds to $UV\rightarrow -1$. By a further transformation
to new null variables $P,Q$, defined through $V=\tan\frac{\pi P}{2}$, 
$U=\tan\frac{\pi Q}{2}$, the global geometry can be represented by a simple
diagram shown in Figure~\ref{fig:global}. Each point in the diagram represents 
an entire $d$-volume parametrized by $\vec{x}$. The global diagram in 
Figure~\ref{fig:global} differs from standard Carter-Penrose conformal diagrams 
in that the boundary at $r\rightarrow\infty$ is not conformally flat.
This is a consequence of the scaling asymmetry between $t$ and $\vec{x}$
and is readily apparent in the globally extended metric (\ref{globalmetric}).

When $z=1$ the auxiliary massive vector field $\mathcal{A}_\mu$ can be
consistently set to zero and the field equations then reduce to those 
of Einstein-Maxwell gravity with a negative cosmological constant. In this 
case, there is a well known exact solution, the AdS-Reissner-Nordström 
black brane, which has a timelike curvature singularity inside an inner and
an outer horizon. 
The interior geometry of the exact $z=2d$ black brane is markedly different
with a null curvature singularity at $r=0$ and no smooth inner horizon. 

\section{Holographic superconductors with asymmetric scaling}
\label{holosc}

We now consider charged black branes with hair. Static spherically
symmetric solutions of the scalar field
equation (\ref{mattereom}) have the asymptotic form 
$\phi(u)\rightarrow c_-(e^{-\Delta_- u}+\ldots)+c_+(e^{-\Delta_+ u}+\ldots)$,
with
\begin{equation}
\Delta_\pm= \frac{d+z}{2} \pm \sqrt{\left( \frac{d+z}{2} \right)^2+m^2},
\label{scalardim}
\end{equation}
while the asymptotic behavior of the electromagnetic field field strength is 
\begin{equation}
\beta(u)\approx \frac{\rho}{r_0^d}e^{-du}+\ldots\, .
\label{betaasymptotic}
\end{equation}
Working at fixed charge density in the dual field theory, we read the radial 
location of the horizon off from the asymptotic behavior of $\beta(u)$ and then
the temperature can be obtained from the numerical solution for 
$f(u)$ and $g(u)$ using (\ref{hawkingtemp}).

In the following we set the scalar mass squared to 
$m^2 =\frac{1}{4}- \left(\frac{d+z}{2}\right)^2$, which is inside the range where 
there is a choice of two boundary theories \cite{Klebanov:1999tb}. 
This choice leads to convenient values, $\Delta_\pm=\frac{d+z}{2}\pm\frac{1}{2}$, 
for the dimensions of the operators $\mathcal{O}_\pm$ that are dual to the scalar field 
in each of the two boundary theories. Non-linear descendants of the leading 
scalar field modes are suppressed by $O(e^{-2u})$ at $u\rightarrow \infty$ and
this choice of mass squared ensures that the first descendant of $\psi_-$ falls off
faster than $\psi_+$.

\begin{figure}[h]
\begin{minipage}[b]{17pc}
\includegraphics[width=17pc]{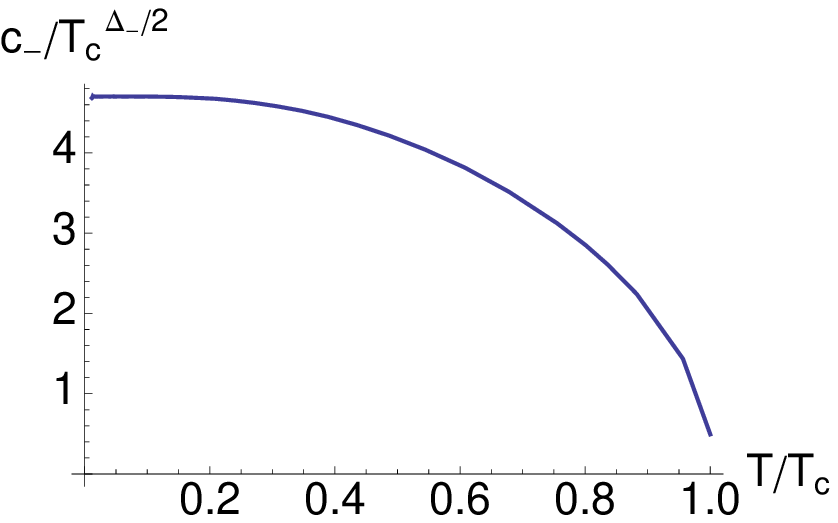}
\caption{\label{fig:condensate}Scalar field condensate in a holographic superconductor 
at $z=d=2$.\hfill\break \ }
\end{minipage}\hspace{2pc}
\begin{minipage}[b]{18pc}
\includegraphics[width=17pc]{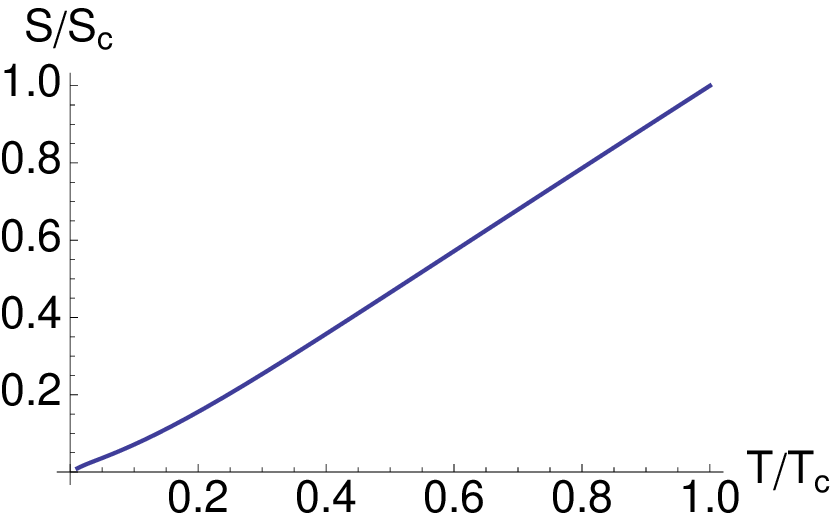}
\caption{\label{fig:zerotentropy}Entropy as a function of temperature 
in the superconducting phase of the holographic superconductor 
in Figure~\ref{fig:condensate}. }
\end{minipage} 
\end{figure}

In order to study holographic superconductivity, we first select some value for 
$d$, $z$ and the
electric charge $q$ carried by the scalar field and then generate numerical black 
brane solutions for a range of initial values $\beta(0)$ and $\phi(0)$. We then 
investigate the asymptotic large $u$ behavior of the scalar hair in
the numerical solutions. A superconducting condensate 
corresponds to either
\begin{equation}
c_+=0,\quad \langle \mathcal{O}_-\rangle =c_- \neq 0, 
\qquad\mathrm{or}\qquad
c_-=0, \quad \langle \mathcal{O}_+\rangle = c_+\neq 0,
\label{czero}
\end{equation}
depending on which of the two boundary theories is being 
considered \cite{Gubser:2008px,Hartnoll:2008vx}. 
We look for a curve in the $\beta(0)$ {\it vs.} $\psi(0)$
plane of initial values at the horizon, for which the corresponding black brane
solution has vanishing $c_+$ ($c_-$), and tabulate the value of  
$c_-$ ($c_+$) along this curve.
The temperature is found from the same numerical solutions {\it via\/} (\ref{hawkingtemp}) 
and (\ref{betaasymptotic}). Figure~\ref{fig:condensate} shows a plot of $c_-$ {\it vs.} $T$
obtained by this procedure for $d=2$, $z=2$, and $q=1$. The results are expressed
in terms of dimensionless ratios that are insensitive to the overall scale (set by 
the charge density $\rho$, which is held fixed at some finite value throughout). 

These results demonstrate that a superconducting condensate can form in systems with 
anisotropic scaling but we have not touched on a number of interesting topics including
the electric conductivity and magnetic properties of these holographic superconductors.

\section{Zero temperature entropy vs. scalar hair}
\label{zerotentropy}
In the absence of charged matter, the charged black branes in our model
have an extremal limit given by 
$\tilde{\rho}\rightarrow \pm\sqrt{2(z^{2}+(d{-}1)z+d^2)}$. It then follows from
$\tilde{\rho}=\rho/r_0^d$ that the radial location
of the horizon $r_0$ has a finite value in the extremal limit for fixed charge
density $\rho$. This in turn means that the entropy density 
$S=r_0^d/4$ remains finite in the zero temperature limit indicating a 
macroscopic groundstate degeneracy. 

This conclusion is radically altered when the system is coupled to a charged 
scalar field. In this case the zero temperature limit is approached in the
condensed phase and the black holes, that are dual to extreme low temperature
states, have scalar hair. It is straightforward to keep track of the black brane
entropy as the temperature is lowered below the critical value for forming
the superconducting condensate. The result for the same $d=2$, $z=2$ 
holographic superconductor as was considered in the previous section is shown 
in Figure~\ref{fig:zerotentropy}. Both the entropy density and the temperature are
normalized to their values at the onset of condensation, $S_c$ and $T_c$ 
respectively.

Although the numerical calculations break down
before absolute zero is reached, the numerical data strongly suggest that
$S$ vanishes in the $T\rightarrow 0$ limit for a generic dynamical critical
exponent $z$, which is consistent with a
non-degenerate ground state in the dual field theory. The corresponding
result for conformal systems with $z=1$ was established in \cite{Horowitz:2009ij}.

\section{Summary}

We have presented an overview of the construction of charged black brane 
solutions in gravity models that realize the anisotropic scaling symmetry 
that is characteristic of many interesting quantum critical points. The motivation
for the study of these models comes from condensed matter theory, 
in particular from two- and three-dimensional systems involving strongly
correlated electrons. The relevance of gravitational models to real world 
condensed matter systems remains highly speculative, but the gravitational
approach continues to produce effects that are intriguingly similar to what is 
seen in experiments. A recent example from our own work \cite{Brynjolfsson:2010rx} 
involves non-Fermi-liquid behavior in the specific heat of anisotropic black 
branes of the type considered in the present paper, which turns out to be
qualitatively similar to the measured specific in certain heavy fermion 
metals near a quantum phase transition \cite{Lohneysen:1994,Stewart:2001zz}.
While much of the work on gravitational modeling of strongly coupled
field theories to date involves asymptotically AdS spacetime and an underlying 
conformal symmetry, it was crucial to the success of this particular application
to have a non-trivial dynamical critical exponent $z>1$.
This provides impetus for further study of gravity models with anisotropic scaling.

\ack

This work was supported in part by the G\"{o}ran Gustafsson foundation, the 
Swedish Research Council (VR), the Icelandic Research Fund, the University of 
Iceland Research Fund, and the Eimskip Research Fund at the University of Iceland.

\end{document}